\documentclass{ws-ijmpd}

\def\bea{\begin{eqnarray}}
\def\beann{\begin{eqnarray*}}
\def\beq{\begin{equation}}
\def\eea{\end{eqnarray}}
\def\eeann{\end{eqnarray*}}
\def\eeq{\end{equation}}
\def\nn{\nonumber}

\begin{document}

%%%%%%%%%%%%%%%%%%%%%%%%%%% To switch off trimmarks %%%%%%%%%%%%%%%%%%%
%

\def\nocropmarks{\vskip5pt\phantom{cropmarks}}

%\let\trimmarks\nocropmarks      %%% Pls. remove the comment sign (%) to switch off the trimmarks

%
%%%%%%%%%%%%%%%%%%%%%%%%%%%%%%%%%%%%%%%%%%%%%%%%%%%%%%

\markboth{M. L. L. da Silva, D. Hadjimichef}
{Glueball-Glueball Interaction in The Context of an Effective Theory}

%%%%%%%%%%%%%%%% Publisher's Area please ignore %%%%%%%%%%%%%%%
%
\catchline{}{}{}
%
%%%%%%%%%%%%%%%%%%%%%%%%%%%%%%%%%%%%%%%%%%%%%%%%%

\title{GLUEBALL-GLUEBALL INTERACTION IN THE CONTEXT\\
OF AN EFFECTIVE THEORY 
%\footnote{For the title, try not to use more than 3 lines.
%Typeset the title in 10 pt roman, uppercase and
%boldface.}
}

\author{\footnotesize MARIO L. L. DA SILVA
%\footnote{E-mail: mariolls@if.ufrgs.br}
}

\address{Instituto de F\'{\i}sica, Universidade Federal do Rio Grande do
  Sul, Av. Bento Gon\c calves, 9500\\
Porto Alegre, Rio Grande do Sul, CEP 91501-970,
Brazil\\
mariolls@if.ufrgs.br
%\footnote{Rio Grande do Sul, Instituto de Fisica, Universidade Federal do Rio Grande do
%  Sul , Av. Bento Goncalves, 9500, CEP 91501-970 Postal Box 15051, Brazil.}
}

\author{DIMITER HADJIMICHEF
%\footnote{E-mail: dimiter@ufpel.edu.br}
}

\address{Instituto de F\'{\i}sica e Matem\'atica, Universidade Federal de
  Pelotas, Campus Universit\'ario,\\
  Pelotas, Rio Grande do Sul, CEP 96010-900, Brazil\\
  dimiter@ufpel.edu.br
}

\maketitle

\pub{Received (received date)}{Revised (revised date)}

\begin{abstract}
  In this work we use a mapping technique to derive in the context of a constituent
  gluon model an effective Hamiltonian that involves explicit gluon degrees of freedom.
  We study glueballs with two gluons using the Fock-Tani formalism.
\end{abstract}


\section{Introduction}	%) A SECTION HEADING

  The gluon self-coupling in QCD implies the existence of bound states of pure gauge
fields known as glueballs. Numerous technical difficulties have so far been present
in our understanding of their properties in experiments, largely because glueball states
can mix strongly with nearby $q\bar{q}$ resonances. However recent experimental and lattice studies
of $0^{++}$, $2^{++}$ and $0^{-+}$ glueballs seem to be convergent. In the present we follow a
different approach by applying the Fock-Tani formalism in order to obtain an effective interaction
between glueballs \cite{annals}. A glueball-glueball cross-section can be obtained and compared with
 usual meson-meson cross-sections.

\section{Fock-Tani Formalism for Glueballs}

The starting point is  the creation operator of a glueball formed by two constituent
gluons
  \begin{eqnarray*}
    G_{\alpha}^{\dagger}=\frac{1}{\sqrt{2}}\Phi_{\alpha}^{\mu \nu}
    a_{\mu}^{\dagger}a_{\nu}^{\dagger}
  \end{eqnarray*}
  where gluons obey the following commutation relations
  \begin{eqnarray*}
    [a_{\mu},a_{\nu}]=0 \,\,\,\, ; \,\,\,\,[a_{\mu},a_{\nu}^{\dagger}]=\delta_{\mu\nu}
  \end{eqnarray*}
  The composite glueball operator satisfy non-canonical commutation relations
  \begin{eqnarray*}
    [G_{\alpha},G_{\beta}]=0\,\,\,\,\,;\,\,\,\, [G_{\alpha},G_{\beta}^{\dagger}]
    =\delta_{\alpha\beta}+\Delta_{\alpha\beta}
  \end{eqnarray*}
  where
  \begin{eqnarray*}
    \delta_{\alpha\beta}=\Phi_{\alpha}^{\star\rho \gamma}\Phi_{\beta}^{\gamma \rho}
\,\,\,\,\,\,;\,\,\,\,\,\,
    \Delta_{\alpha\beta}=2\Phi_{\alpha}^{\star\mu
      \gamma}\Phi_{\beta}^{\gamma \rho}a_{\rho}^{\dagger}a_{\mu}
  \end{eqnarray*}
  The Fock-Tani formalism introduces ``ideal particles'' which obey canonical relations,
in our case they are ideal glueballs
  \begin{eqnarray*}
    [g_{\alpha},g_{\beta}]=0\,\,\,\,\,;\,\,\,\, [g_{\alpha},g_{\beta}^{\dagger}]
    =\delta_{\alpha\beta}
  \end{eqnarray*}
  This way one can transform the composite glueball state $|\alpha\rangle$  into an ideal state
$|\alpha\,)$ by
  \begin{eqnarray*}
 |\alpha\,)=   U^{-1}(-\frac{\pi}{2})\,G_{\alpha}^{\dagger}
\,|0\rangle=g_{\alpha}^{\dagger}\,| 0\rangle
  \end{eqnarray*}
  where $ U=\exp({tF})$ and $F$ is the generator of the glueball transformation given by
  \begin{eqnarray}
    F=g_{\alpha}^{\dagger}\tilde{G}_{\alpha}-\tilde{G}_{\alpha}^{\dagger}g_{\alpha}
    \label{F}
  \end{eqnarray}
  with
  \begin{eqnarray*}
    \tilde{G}_{\alpha}=G_{\alpha}-\frac{1}{2}\Delta_{\alpha\beta}G_{\beta}
    -\frac{1}{2}G_{\beta}^{\dagger}[\Delta_{\beta\gamma},G_{\alpha}]G_{\gamma}
  \end{eqnarray*}
  In order to obtain the effective glueball-glueball potential one has to use
  (\ref{F}) in a set of Heisenberg-like equations for the basic operators $g,\tilde{G},a$
  \begin{eqnarray*}
    \frac{dg_{\alpha}(t)}{dt}=[g_{\alpha},F]=\tilde{G}_{\alpha}\,\,\,\,\,;\,\,\,\,\,
    \frac{d\tilde{G}_{\alpha}(t)}{dt}=[\tilde{G}_{\alpha}(t),F]=-g_{\alpha}\,.
  \end{eqnarray*}
  The simplicity of these equations are not present in the equations for $a$
  \begin{eqnarray*}
    \frac{da_{\mu}(t)}{dt}=[a_{\mu},F]=&-&\sqrt{2}\Phi_{\beta}^{\mu\nu}a_{\nu}^{\dagger}g_{\beta}
    +\frac{\sqrt{2}}{2}\Phi_{\beta}^{\mu\nu}a_{\nu}^{\dagger}\Delta_{\beta\alpha}g_{\beta}\\
    &+&\Phi_{\alpha}^{\star\mu\gamma}\Phi_{\beta}^{\gamma\mu^{'}}
    (G_{\beta}^{\dagger}a_{\mu^{'}}g_{\beta}-g_{\beta}^{\dagger}a_{\mu^{'}}G_{\beta})\\
    &-&\sqrt{2}(\Phi_{\alpha}^{\mu\rho^{'}}\Phi_{\rho}^{\mu^{'}\gamma^{'}}
    \Phi_{\gamma}^{\star\gamma^{'}\rho^{'}}+\Phi_{\alpha}^{\mu^{'}\rho^{'}}\Phi_{\rho}^{\mu\gamma^{'}}
    \Phi_{\gamma}^{\star\gamma^{'}\rho^{'}})G_{\gamma}^{\dagger}a_{\mu^{'}}^{\dagger}
    G_{\beta}g_{\beta}
  \end{eqnarray*}
  The solution for these equation can be found order by order in the wavefunctions. So, for zero order
  one has $a_{\mu}^{(0)}=a_{\mu}$
  \begin{eqnarray*}
    g_{\alpha}^{(0)}(t)=G_{\alpha}\sin{t}+g_{\alpha}\cos{t}\,\,\,\,;\,\,\,\,G_{\beta}^{(0)}(t)
    =G_{\beta}\cos{t}-g_{\beta}\sin{t}
  \end{eqnarray*}
  In the first order $g_{\alpha}^{(1)}=0,\,\,\,G_{\beta}^{(1)}=0$ and
  \begin{eqnarray*}
    a_{\mu}^{(1)}(t)=\sqrt{2}\Phi_{\beta}^{\mu\nu}a_{\nu}^{\dagger}[G_{\beta}^{(0)}-G_{\beta}]
  \end{eqnarray*}
  In the second order we found
  \begin{eqnarray*}
    a_{\mu}^{(2)}(t)=-2\Phi_{\alpha}^{\star\mu\gamma}\Phi_{\beta}^{\gamma\mu^{'}}
    G_{\beta}^{\dagger}a_{\mu^{'}}G_{\alpha}^{(0)}+\Phi_{\alpha}^{\star\mu\gamma}
    \Phi_{\beta}^{\gamma\mu^{'}}G_{\beta}^{\dagger}a_{\mu^{'}}G_{\alpha}\nonumber
    +\Phi_{\alpha}^{\star\mu\gamma}\Phi_{\beta}^{\gamma\mu^{'}}G_{\beta}^{\dagger(0)}
    a_{\mu^{'}}G_{\alpha}^{(0)}
  \end{eqnarray*}
To obtain the third order $a_{\mu}^{(3)}(t)$ is straightforward and shall be presented elsewhere.
%  \begin{eqnarray*}
%    a_{\mu}^{(3)}(t)=&+&\sqrt{2}\Phi_{\alpha}^{\mu\nu}\Phi_{\beta}^{\star\sigma\nu}
%    \Phi_{\gamma}^{\sigma\tau}(G_{\beta}^{\dagger}a_{\tau}^{\dagger}G_{\gamma}G_{\alpha}^{(0)}
%    -G_{\beta}^{\dagger(0)}a_{\tau}^{\dagger}G_{\gamma}G_{\alpha}^{(0)}\nonumber\\
%    &+&G_{\beta}^{\dagger(0)}a_{\tau}^{\dagger}G_{\gamma}^{(0)}G_{\alpha}^{(0)}
%    -G_{\beta}^{\dagger}a_{\tau}^{\dagger}G_{\gamma}G_{\alpha})\nonumber\\
%    &-&\frac{\sqrt{2}}{2}\Phi_{\alpha}^{\mu\nu}a_{\nu}^{\dagger}\Delta_{\alpha\gamma}
%    [(\cos{t}-2)G_{\gamma}+G_{\gamma}^{(0)}]
%  \end{eqnarray*}
  The glueball-glueball potential can be obtained applying in a standard way the Fock-Tani
  transformed operators to the microscopic Hamiltonian
  \begin{eqnarray*}
    {\cal{H}}(\mu\nu;\sigma\rho)=T_{\rm aa}(\mu)a_{\mu}^{\dagger}a_{\mu}+\frac{1}{2}
    V_{\rm aa}(\mu\nu;\sigma\rho)a_{\mu}^{\dagger}a_{\nu}^{\dagger}a_{\rho}a_{\sigma}
  \end{eqnarray*}
  where one obtains for the glueball-gluball potential $V_{gg}$
  \bea
    V_{gg}=\sum_{i=1}^{4}V_{i}(\alpha\gamma;\delta\beta)g_{\alpha}^{\dagger}
    g_{\gamma}^{\dagger}g_{\delta}g_{\beta}
\label{v_gg}
  \eea
  and
  \begin{eqnarray*}
    &V_{1}(\alpha\gamma;\delta\beta)=2V_{aa}(\mu\nu;\sigma\rho)\Phi_{\alpha}^{\star\mu\tau}
    \Phi_{\gamma}^{\star\nu\xi}\Phi_{\delta}^{\rho\xi}\Phi_{\beta}^{\sigma\tau}\\
    &V_{2}(\alpha\gamma;\delta\beta)=2V_{aa}(\mu\nu;\sigma\rho)\Phi_{\alpha}^{\star\mu\tau}
    \Phi_{\gamma}^{\star\nu\xi}\Phi_{\delta}^{\rho\tau}\Phi_{\beta}^{\sigma\xi}\\
    &V_{3}(\alpha\gamma;\delta\beta)=V_{aa}(\mu\nu;\sigma\rho)\Phi_{\alpha}^{\star\mu\nu}
    \Phi_{\gamma}^{\star\lambda\xi}\Phi_{\delta}^{\sigma\lambda}\Phi_{\beta}^{\rho\xi}\\
    &V_{4}(\alpha\gamma;\delta\beta)=V_{aa}(\mu\nu;\sigma\rho)\Phi_{\alpha}^{\star\mu\xi}
    \Phi_{\gamma}^{\star\nu\lambda}\Phi_{\delta}^{\lambda\xi}\Phi_{\beta}^{\rho\sigma}\,.
  \end{eqnarray*}
\vspace{-1.0cm}
\begin{center}
\begin{figure*}[htb]
\epsfxsize=25pc \hspace{1cm} \epsfbox{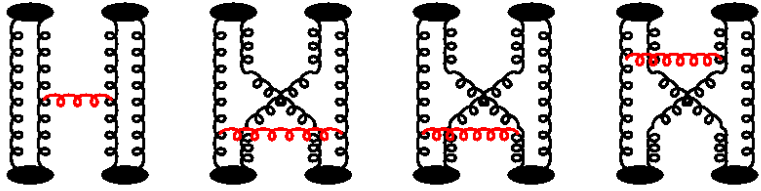} \caption{Diagrams
representing the scattering amplitude $h_{fi}$ for
glueball-glueball interaction with constituent gluon interchange.}
\label{diagram}
\end{figure*}
\end{center}
\vspace{-1.0cm}
The next step is to obtain the  scattering $T$-matrix from Eq. (\ref{v_gg})
\beann
T(\alpha\beta;\gamma\delta)=(\alpha\beta|V_{gg}|\gamma\delta)\,.
\eeann
Due to translational invariance, the $T$-matrix element is written as a momentum conservation
delta-function, times a Born-order matrix element, $h_{fi}$:
$
T(\alpha\beta;\gamma\delta)=\delta^{(3)}( \vec{P}_{f}-\vec{P}_{i})\,h_{fi}
$,
where $\vec{P}_{f}$ and  $\vec{P}_{i}$
are the final and initial momenta of the two-glueball system.  This result can be used
in order to evaluate  the glueball-glueball scattering cross-section
\bea
\sigma_{gg} =\frac{4\pi ^{5}\,s}{s-4M_{G}^{2}}
\int_{-(s-4M_{G}^{2})}^{0}\,dt\,|h_{fi}|^{2}
\label{cross}
\eea
where $M_G$ is the glueball mass, $s$ and $t$ are the Mandelstam variables.

\section{The Constituent Gluon Model}

 On theoretical grounds, a 
simple potential model with massive constituent gluons, namely the model of Cornwall and 
Soni \cite{cs1},\cite{cs2} has been  studied \cite{cs3},\cite{cs4} and the results are 
consistent with lattice and experiment. In the conventional quark model a 
$0^{++}$ state is considered as $q\bar{q}$ bound state.  The $0^{++}$ resonance is a isospin 
zero state so, in principal, it can be either represented as a  $q\bar{q}$  bound state, a 
glueball, or a mixture. In particular there is growing evidence in the direction of large 
$s\bar{s}$ content with some mixture with the glue sector. 
It turns out that this resonance is an interesting system, in the theoretical point of view, 
where one can compare models. In the present work we consider 
 two possibilities for $0^{++}$: ({\it i}) a as pure $s\bar{s}$ and calculate,
in the context of a quark interchange picture, the cross-section; ({\it ii} ) as a glueball
where a new calculation for this cross-section is made, in the context of
the constituent gluon model, with gluon interchange. The potential $V_{\rm aa}$ is determined
in the Cornwall and Soni constituent gluon model \cite{cs1}
\bea
V_{\rm aa}(r)= \frac{1}{3}\, f^{ace}f^{bde}   \left[\,V_{2g}^{OGEP}(r)
+V_{S}(r)
\right]
\label{mod24}
\eea
where
\bea
V_{2g}^{OGEP}(r)=   -\lambda\,
 \left[ \,\omega_{1}\frac{e^{-m r}}{r}
+\omega_{2}\frac{\pi}{m^2}\,D(r)\,
\right]\,\,\,\,\,\,,\,\,\,\,\,\,
V_{S}(r)= 2m\,(1-e^{-\beta \,m\,r})
\label{mod24a}
\eea
and
\bea
D(r)=\frac{k^{3}m^{3}}{\pi^{3/2}}\,e^{-k^{2}m^{2}r^{2}}
\,\,\,\,\,\,,\,\,\,\,\,\,
\lambda = \frac{3\,g^{2}}{4\pi} 
\,\,\,\,\,\,,\,\,\,\,\,\,
\omega_{1} = \frac{1}{4}+\frac{1}{3}\vec{S}^2 
\,\,\,\,\,\,,\,\,\,\,\,\,
\omega_{2} = 1-\frac{5}{6}\vec{S}^2  .
\label{mod24b}
\eea
The parameters $\lambda$, $m$, $k$ and $\beta$ assume known values\cite{cs2,cs3,cs4}
while  the wave function $\Phi^{\mu\nu}_{\alpha}$ is given in \cite{sergio}.
The glueball-glueball scattering amplitude $h_{fi}$  is given by
\bea
  h_{fi}(s,t) &=& \frac{3}{8}\,R_{0}(s)
\sum_{i=1}^{6}\,R_{i}(s,t)
\label{hfi-st}
\eea
where
\bea
R_{0}&=& \frac{4}{(2\pi)^{3/2}b^3}
  \exp\left [-\frac{1}{2b^2} \left (\frac{s}{4} -M_{G}^2 \right)  \right]\nn\\
R_{1}&=&\frac{\lambda \omega_{1}^{(2)}\, 4\,\sqrt{2\pi}}{3     }  \,
  \int_{0}^{\infty} dq\,\frac{q^2}{q^2 +m^{2}}
  \exp\left(-\frac{q^2}{2b^2}\right) 
\left[
{\mathcal{J}}_0  \left(\frac{q\sqrt{t}}{2b^2}\right) +
{\mathcal{J}}_0  \left(\frac{q\sqrt{u}}{2b^2}\right)
\right] \nn\\
R_{2}&=&
\frac{  \lambda \omega_{2}^{(2)}     2\sqrt{2} \pi b^3 k^3 m}{3(b^2
  +2k^2 m^{2})^{3/2}} 
\left[
\exp \left( -\frac{t k^2 m^2}{4(b^4 +2b^2 k^2 m^{2})}\right)
%\right.
%\nn\\
%&&
%+\left.
+\exp \left( -\frac{u k^2 m^2}{4(b^4 +2b^2 k^2 m^{2})}\right)
\right]\nn\\
R_{3} &=&
\frac{32\sqrt{2\pi}}{3} \int_0^{\infty} dq\,
  \frac{q^2 \beta  m^{2}}{(q^{2} +\beta^{2}m^{2})^{2}} \exp
  \left(-\frac{q^2}{2b^2}\right)
\left[
 {\mathcal{J}}_0 \left(\frac{q\sqrt{t}}{2b^2}  \right)+
 {\mathcal{J}}_0 \left(\frac{q\sqrt{u}}{2b^2}  \right)
\right]\nn\\
R_{4}&=&
-\frac{\lambda \omega_{1}^{(3)}}{3}\,\,
  \frac{16\,\sqrt{2\pi}\,b^2}{\sqrt{\frac{s}{4}-M_{G}^2}}
 \int_{0}^{\infty} dq\,\frac{q}{q^2 +m^{2}}
  \exp\left(-\frac{3q^2}{8b^2}\right)
  \sinh\left(\frac{q}{2b^2}\sqrt{\frac{s}{4}-M_{G}^2}\, \right) \nn\\
R_{5}&=&-\frac{\lambda \omega_{2}^{(3)}\,}{3}\, \frac{16\pi b^3 k^3 m}{(2b^2
  +3k^2 m^{2})^{3/2}} \exp \left[ -\frac{k^2 m^2 \left( \frac{s}{4}
  -M_{G}^2\right)}{2(2b^4 +3b^2 k^2 m^{2})}\right]\nn\\
R_{6}&=&-\frac{128 \sqrt{2\pi}
  b^2}{3\sqrt{\frac{s}{4} -M_{G}^2}} \int_0^{\infty} dq\,
  \frac{q \beta  m^{2}}{(q^{2} +\beta^{2}m^{2})^{2}}
  \exp \left(\frac{3q^2}{8b^2}\right)  \sinh \left(\frac{q}{2b^2}
  \sqrt{\frac{s}{4}-M_{G}^2}\, \right)
\label{R}
\eea
here $b=\frac{\sqrt{3}}{\sqrt{2}\,r_{0} } $ where $r_{0}$ is the glueball's {\it rms} radius  
and  ${\cal J}_{0}(x)=\sin x /x $. In (\ref{R}) one finds the following notation $\omega^{(i)}_{1}$
and $\omega^{(i)}_{2}$, where the index $i$ correponds to the number of the evaluated diagram
in figure (1)
The cross-section is obtained inserting (\ref{hfi-st})
in (\ref{cross}). From reference \cite{sergio} one obtains the corresponding cross-section
for a $0^{++}$ meson with a $s\bar{s}$ content
\bea
\sigma_{fi}^{s\bar{s}}=\frac{4\pi\alpha_{s}^{2}s}{81m_{q}^{4}}
\left[\frac{4b^2\left(1-e^{-\frac{\xi}{4b^2}}
\right)}{\xi}+\frac{128}{27}e^{-\frac{\xi}{6b^2}}+e^{-\frac{\xi}{8b^2}}
+\frac{64}{3\sqrt{3}}\frac{4b^2}{\xi}\left(e^{-\frac{\xi}{12b^2}}
-e^{-\frac{5\xi}{24b^2}}
\right)\right]\;\nonumber
\eea
with $\xi=s-4M_{G}^2$. The comparison between the cross-sections in the glueball picture and the 
quark picture for the $0^{++}$ meson is given in figure (2).

\begin{figure}[htb]
\begin{center}
{\epsfig{file=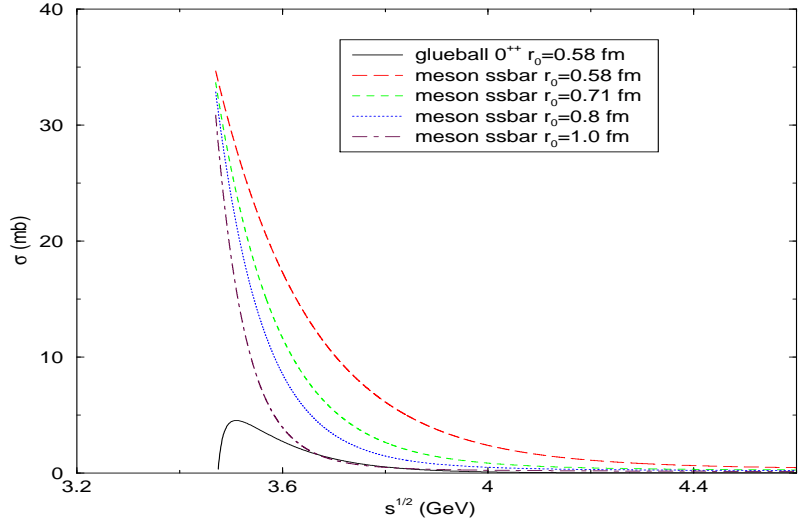, height=300.00pt, width=200.00pt,angle=-90}}
\end{center}
\caption{Cross-section comparison for $0^{++}$ with the following parameters
 $\beta=0.1$, $\lambda = 1.8$, $k=0.21$, gluon mass $m=0.6$GeV. The  $s\bar{s}$ quark 
model parameters: $m_{q}=0.55$ GeV, $\alpha_{s}=0.6$.
}
\label{graf_5}
\end{figure}
\vspace{-0.5cm}

\section{Conclusions}

In this work we have extended the Fock-Tani Formalism to a hadronic model in which the bound state is composed by bosons.
The Cornwall-Soni constituent gluon model has been successful in describing low mass glueballs, in particular
the $0^{++}$ resonance, which is a isospin  zero state. This state  can be either represented as a  $q\bar{q}$  
bound state, a  glueball, or a mixture. In the present work we have considered
 two possibilities for $0^{++}$: a as pure $s\bar{s}$ and as a glueball. A comparison of the cross-sections
reveals that a quark composition for the $0^{++}$ implies in a larger {\it rms} radius than in the
constituent gluon picture. This could represent a criterion for distinguishing between pictures.

%\vspace{-0.5cm}

\section*{Acknowledgements}

The author (M.L.L.S.)  was supported by Conselho Nacional de Desenvolvimento Cient\'{\i}fico e Tecnol\'ogico (CNPq).

\end{document}